\documentclass[prc,twocolumn,letterpaper,10pt,twoside,tightenlines,nofootinbib,showpacs]{revtex4}
\usepackage{amsmath,amsfonts,amssymb,latexsym}
\usepackage{graphicx}
\usepackage[sort&compress]{natbib}
\begin{document}

\newcommand{\eg}{{\it e.g.}}
\newcommand{\etal}{{\it et. al.}}
\newcommand{\ie}{{\it i.e.}}
\newcommand{\be}{\begin{equation}}
\newcommand{\ee}{\end{equation}}
\newcommand{\bea}{\begin{eqnarray}}
\newcommand{\eea}{\end{eqnarray}}
\newcommand{\bef}{\begin{figure}}
\newcommand{\eef}{\end{figure}}
\newcommand{\bce}{\begin{center}}
\newcommand{\ece}{\end{center}}

\newcommand{\dd}{\text{d}}
\newcommand{\ii}{\text{i}}
\newcommand{\lsim}{\lesssim}
\newcommand{\gsim}{\gtrsim}

\title{$\mathbf{D_s}$-Meson as Quantitative Probe of Diffusion and
  Hadronization in Nuclear Collisions}

\author{Min~He, Rainer~J.~Fries and Ralf~Rapp}
\affiliation{Cyclotron Institute and Department of Physics and
Astronomy,
       Texas A\&M University, College Station, Texas 77843-3366, U.S.A.}

\date{\today}

\begin{abstract}
The modifications of $D_s$-meson spectra in ultrarelativistic heavy-ion
collisions are identified as a quantitative probe of key
properties of the hot nuclear medium. This is enabled by the unique
valence-quark content of the $D_s$=$c\bar{s}$ which couples the well-known
strangeness enhancement with the collective-flow pattern of primordially
produced charm quarks. We employ a consistent strong-coupling treatment
with hydrodynamic bulk evolution and nonperturbative $T$-matrix interactions
for both heavy-quark diffusion and hadronization in the Quark-Gluon Plasma
(QGP). A large enhancement of the $D_s$ nuclear modification factor
($R_{AA}$) at RHIC is predicted, with a remarkable maximum of $\sim$1.5-1.8 at
transverse momenta around 2~GeV/$c$. We show this to be a direct consequence of
the strong coupling of the heavy quarks to the QGP and their hadronization
via coalescence with strange quarks. We furthermore introduce the effects of
diffusion in the hadronic phase and suggest that an increase of the
$D$-meson elliptic flow compared to the $D_s$ can disentangle the
transport properties of hadronic and QGP liquids.
\end{abstract}


\maketitle

The properties of nuclear matter at high temperatures and densities
are under intense investigation. Numerical lattice simulations of
Quantum Chromodynamics (QCD) predict the formation of a deconfined
Quark-Gluon Plasma (QGP) at a transition temperature of
$T_c\simeq170$\,MeV~\cite{Aoki:2006we,Bazavov:2011nk}. The QGP presumably
filled the early universe during the first few microseconds of its
existence. Ultra-relativistic heavy-ion collisions (URHICs) at BNL's
Relativistic Heavy Ion Collider (RHIC) and at CERN's Large Hadron Collider
(LHC) are aimed at producing the QGP and characterizing its properties.
Thus far, the experimental results point to the creation of a
color-opaque, strongly coupled and almost ideal liquid with initial
temperatures well above $T_c$~\cite{Adams:2005dq,Aamodt:2010pa}.
However, this assessment remains
qualitative at present~\cite{Gyulassy:2004zy,Shuryak:2008eq}. A
quantification of the physical properties of the strongly-coupled QGP
in terms of its microscopic interactions calls for novel probes and
observables.

A special role in these efforts is played by heavy quarks (charm and bottom,
$Q$=$c$,$b$). Their masses ($m_{c,b}\simeq1.3,4.5$\,GeV) are significantly
larger than the typical temperatures of a few hundred MeV reached at RHIC and
LHC.
Therefore, their production is essentially restricted to primordial
nucleon-nucleon collisions~\cite{Adler:2004ta}, and their thermal relaxation
during the subsequent evolution of the medium is delayed relative to light
quarks by a factor of $\sim m_Q/T\approx$\,5-20~\cite{Svetitsky:1987gq,vanHees:2004gq,Moore:2004tg,Mustafa:2004dr,Rapp:2009my}.
This renders the thermalization time of heavy quarks comparable to the typical
lifetime of the QGP phase in URHICs, and thus makes them excellent probes.
Current heavy-quark (HQ) observables, mostly in terms of pertinent semi-leptonic
electron-decay ($e^\pm$) spectra~\cite{Abelev:2006db,Adare:2006nq,Adare:2010de},
indicate a surprising degree of thermalization and collectivity,
especially through a large elliptic flow~\cite{Adare:2006nq,Adare:2010de}.
First data on $D$-meson spectra confirm this behavior~\cite{Zhang:2011uva}.

The thermalization and collective flow imparted on heavy quarks by
the ambient expanding QGP serves as a quantitative measure of their
coupling to thermalized light quarks and gluons. Current experimental
results call for an interaction strength that goes well beyond
perturbative QCD (pQCD) with realistic strong coupling constants,
$\alpha_s\simeq0.3$~\cite{vanHees:2005wb,Wicks:2005gt,Adil:2006ra,vanHees:2007me,Gossiaux:2008jv,Das:2009vy,
Alberico:2011zy,Uphoff:2011ad}. Calculations based on nonperturbative heavy-light
quark interactions, followed by coalescence into $D$
mesons~\cite{vanHees:2005wb,vanHees:2007me}, describe the observed
modifications of $e^\pm$ spectra reasonably well, up to transverse
momenta of $p_t^e\simeq5$\,GeV/$c$. However, since the $e^\pm$ spectra
are a superposition of
charm and bottom contributions, an enhanced discrimination power is
desirable. In this Letter we argue that accurate measurements of
$D_s$ mesons considerably improve our ability to constrain the
interactions of heavy flavor in medium: by directly comparing $D$
and $D_s$ spectra and elliptic flow one can disentangle recombination
effects in hadronization, as well as the charm coupling to the QGP and
hadronic matter.

One of our key points here is to test the dual role of the resonant
HQ interactions with light quarks -- figuring into both diffusion
and hadronization -- by utilizing the well-established enhancement
of strangeness production in URHICs. The latter is among the
earliest suggested signatures of QGP formation~\cite{Koch:1986ud}.
It manifests itself as a $\sim$50\% increase of
strange-to-light hadron ratios in central nucleus-nucleus (AA)
relative to $pp$ collisions (e.g., in the $K/\pi$
ratio)~\cite{Abelev:2008ez,Adler:2003cb},
and can be understood as a strangeness equilibration within the
statistical hadronization model~\cite{BraunMunzinger:2003zd}. Thus,
if resonant HQ interactions and recombination with thermalized light
quarks into $D$-mesons are key to explaining HQ observables, heavy
quarks must also couple to the equilibrated strangeness
content of the QGP. In contrast to earlier predictions of inclusive
$D_s$ yields in URHICs~\cite{Andronic:2003zv,Kuznetsova:2006bh}, we
here perform a full dynamical treatment of charm diffusion and
hadronization over a large range in transverse momentum ($p_T$),
encompassing thermal and kinetic regimes. Most notably, we predict
the nuclear modification factor ($R_{AA}$) of the $D_s$ to
significantly exceed one, which would be the first of its kind for
a meson at collider energies.
The main, dynamical, information is, however, encoded in its
$p_T$-dependence, which allows us to scrutinize both diffusion and
hadronization effects, especially in comparison to $D$ mesons.

The $D_s$ and $D$ observables encode yet another aspect which has not
been evaluated in the HQ context to date: the role of the hadronic
phase. Following the common, empirically supported notion that
multistrange hadrons (e.g., $\phi$ and $\Omega^-$) decouple close to
the hadronization transition~\cite{Abelev:2008aa,He:2010vw}, the same should
apply to the $D_s$. Hence, after evaluating the effects of hadronic
diffusion on the $D$-meson, a comparison of $D$ and $D_s$ observables
enables a quantitative
assessment of the hadronic transport coefficient. Since the HQ
transport coefficient is believed to be closely related to the
widely discussed viscosity-to-entropy ratio of QCD matter
($\eta/s$)~\cite{Rapp:2009my}, disentangling the former for hadronic
and QGP phases would allow us to quantify the temperature
dependence of $\eta/s$.

\begin{figure}[!t]
\includegraphics[width=\columnwidth]{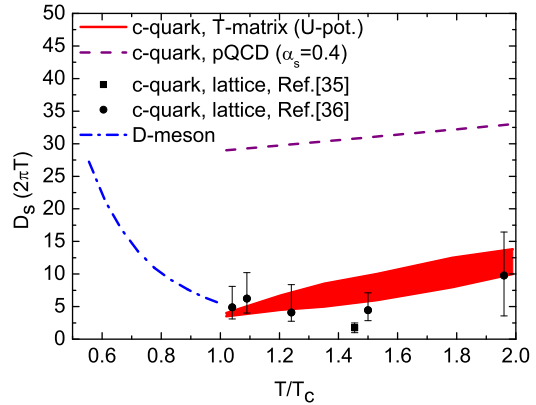}
\caption{(Color online) The spatial diffusion coefficients $\mathcal{D}_s$
(in units of the thermal wavelength, $1/(2\pi T)$) vs.~temperature (in
units of $T_c$) for charm quarks using $T$-matrix
interactions in the QGP (red band) and $D$ mesons using effective
lagrangians in hadronic matter (blue solid line), compared to pQCD
(purple dashed line) and quenched lQCD (data
points)~\cite{Ding:2011hr,Banerjee:2011ra}.}
\label{fig_Ds}
\end{figure}

Let us start by describing our calculations of charm diffusion in
medium. At the temperatures of interest in URHICs, $T$$\lsim$400\,MeV,
HQ kinetics can be approximated by Brownian motion, with a
thermal relaxation rate $A(p,T)$ derived from a Fokker-Planck
equation~\cite{Svetitsky:1987gq,vanHees:2004gq,Moore:2004tg,vanHees:2005wb,vanHees:2007me,Alberico:2011zy}.
The key input to calculate $A$ are HQ scattering amplitudes; at low- and
intermediate momenta, these are governed by elastic interactions of
potential type. In the QGP, we employ thermodynamic $T$-matrix
calculations~\cite{vanHees:2007me,Riek:2010fk} with input potentials
approximated by HQ internal energies computed in thermal lattice QCD
(lQCD).  The resulting $T$-matrices exhibit resonant states close to
threshold in mesonic and diquark channels close to $T_c$.
They accelerate thermal HQ relaxation by a factor of $\sim$3-5 over
leading-order pQCD~\cite{Riek:2010fk}. In the hadronic phase, the thermal
relaxation rate of $D$ mesons is obtained~\cite{He:2011yi} from
elastic scattering amplitudes based on effective hadronic Lagrangians
constrained by vacuum spectroscopy.
Figure~\ref{fig_Ds} summarizes our input for the spatial diffusion
coefficient of charm in matter, defined by the Einstein relation as
$\mathcal{D}_s=T/[m_{c,D}A(p=0,T)]$. Coming from high $T$ we observe
a steady decrease in the QGP (due to an increased interaction strength
in the lQCD potentials), followed by an increase in hadronic matter.
The calculations support a continuous evolution through the transition
region, with a minimum of $\mathcal{D}_s\simeq(4{\rm -}5)/2\pi T$ around
$T_c$ (translating into an estimate of $\eta/s\sim(2{\rm -}4)/4\pi$).
Also note that our results in the QGP are comparable to recent quenched
lQCD estimates~\cite{Ding:2011hr,Banerjee:2011ra}, but far below the
pQCD values.

The Brownian motion of charm is implemented into URHICs via
relativistic Langevin simulations, with the space-time evolution of the medium
approximated by boost-invariant ideal hydrodynamics. To this end
we employ our recent tune~\cite{He:2011zx} of the AZHYDRO
code~\cite{Kolb:2003dz}, optimized to describe bulk and multistrange
hadron spectra and elliptic flow in Au-Au
collisions RHIC. It utilizes a state-of-the-art lQCD equation of
state~\cite{Bazavov:2011nk,Borsanyi:2010cj} with pseudo-critical
deconfinement temperature of $T_c$=170\,MeV, and a subsequent
hadron-resonance-gas phase with chemical freezeout at $T_{\rm
ch}$=160\,MeV to account for the observed hadron ratios. A compact
initial spatial profile with pre-equilibrium flow permits a
simultaneous fit of multistrange- and bulk-hadron data at chemical
and thermal ($T_{\rm fo}$$\simeq$110\,MeV) freezeout, respectively.
The initial HQ distributions are taken from a PYTHIA tune to $e^\pm$
spectra in $pp$ and $d$Au collisions~\cite{vanHees:2005wb}. The
Cronin effect in nuclear collisions is accounted for via a Gaussian
transverse-momentum broadening with $\langle k_T^2 \rangle=0.6~{\rm
GeV^2}$, estimated from recent PHENIX $e^\pm$ spectra in
$d$Au~\cite{durham}.



After diffusion through the QGP charm-quark distributions are
converted into charmed hadrons. We accomplish this by applying
resonance recombination~\cite{Ravagli:2007xx} with thermal light and
strange quarks into $D$ and $D_s$ mesons on the hydro hypersurface
at $T_c$~\cite{He:2011qa}. Remaining $c$-quarks are treated with
$\delta$-function fragmentation (as used in the fits to $pp$ data).
For a reliable coalescence dynamics at low and intermediate $p_T$,
it is crucial that the formulation of the resonance recombination
model (RRM) via a Boltzmann equation~\cite{Ravagli:2007xx} yields
the long-time limit of thermal equilibrium. We have verified this in
the present case with the full space-momentum correlations as given
by the hydrodynamic flow field~\cite{He:2011qa}. The coalescence
probabilities are estimated via $P_{\rm coal}(p)\simeq\Delta\tau_{\rm
res}\Gamma_c^{\rm res}(p)$, with the charm-quark reaction rate,
$\Gamma_c^{\rm res}(p)$ (as given by the heavy-light $T$ matrix),
and a time duration $\Delta\tau_{\rm res}$ characterizing one
generation of $D$ and $D_s$ resonance formation~\cite{He:2011qa}.
With $\Gamma_c^{\rm res}(0)$$\simeq$0.2\,GeV~\cite{Riek:2010py} and
$\Delta\tau_{\rm res}$$\approx$1\,fm/$c$,
we assume a recombination probability of one at vanishing
charm-quark momentum, decreasing thereafter as determined by the
dynamics of the RRM expression~\cite{Ravagli:2007xx}. The latter is
evaluated with $m_c$=1.7\,GeV, $m_{u,d}$=0.3\,GeV, $m_s$=0.4\,GeV
and $m_D$=2.1\,GeV, $m_{D_s}$=2.2\,GeV with
$\Gamma_{D,D_s}=0.2$\,GeV, approximately representing the in-medium
values of the $T$-matrices~\cite{Riek:2010fk} during the
hadronization window. Since HQ resonant scattering is underlying
both diffusion and hadronization interactions, there is, in principle,
some overlap between the two (this does not apply to the non-resonant
parts of the interactions). To characterize this uncertainty, we will
study a scenario where diffusion interactions in the QGP are completely
switched off for about 1\,fm/$c$ prior to $T_c$, corresponding
to a temperature window of 180-170\,MeV.
The Langevin simulation resumes with hadronic diffusion of the
combined coalescence+fragmentation distribution for
$D$-mesons for $T{\rm <}T_c$ until hydrodynamic freezeout at
$T_{\rm fo}$=110\,MeV, while the $D_s$-meson distribution is
frozen at $T_c$.

It remains to determine absolute magnitude of the coalescence
contribution to the $D$ and $D_s$ yields in AA collisions.
In $pp$ collisions we assume fragmentation only with hadronization
fractions from recent PYTHIA simulations~\cite{MartinezGarcia:2007hf},
i.e., $D$/$c$=82\%
and $D_s$/$c$=11\%, including feed-down from excited states (here,
$D$$\equiv$$D^+$+$D^0$). This gives $D_s/D$=0.134 in $pp$, in line
with CDF data in
$p\overline{p}$($\sqrt{s}$=1.96\,TeV)~\cite{Acosta:2003ax,akiba} and
the value used in a recent PHENIX analysis~\cite{Adare:2010de}.
Since our coalescence contribution is evaluated using thermalized
light and strange quarks within RRM, the logical choice for the
pertinent $D_s/D$ ratio are thermal weights,
which we adopt from the statistical hadronization model
(including feeddown from excited states)~\cite{Andronic:2007zu}, with
an additional strangeness fugacity, $\gamma_s$=0.85~\cite{Abelev:2008ez},
for consistency with the hadronic equation of state in our hydro
evolution~\cite{He:2011zx}.
Then, upon combining coalescence and fragmentation with the probabilities
elaborated above, we obtain $D$/$c$=75\% and $D_s$/$c$=15\%, or
$D_s/D$=0.20, for Au+Au collisions at $\sqrt{s_{\rm NN}}$=200\,GeV,
hence obtaining an enhancement of the $D_s/D$ ratio of
$\sim$50\% over the value in $pp$.

\begin{figure}[!t]
\includegraphics[width=\columnwidth]{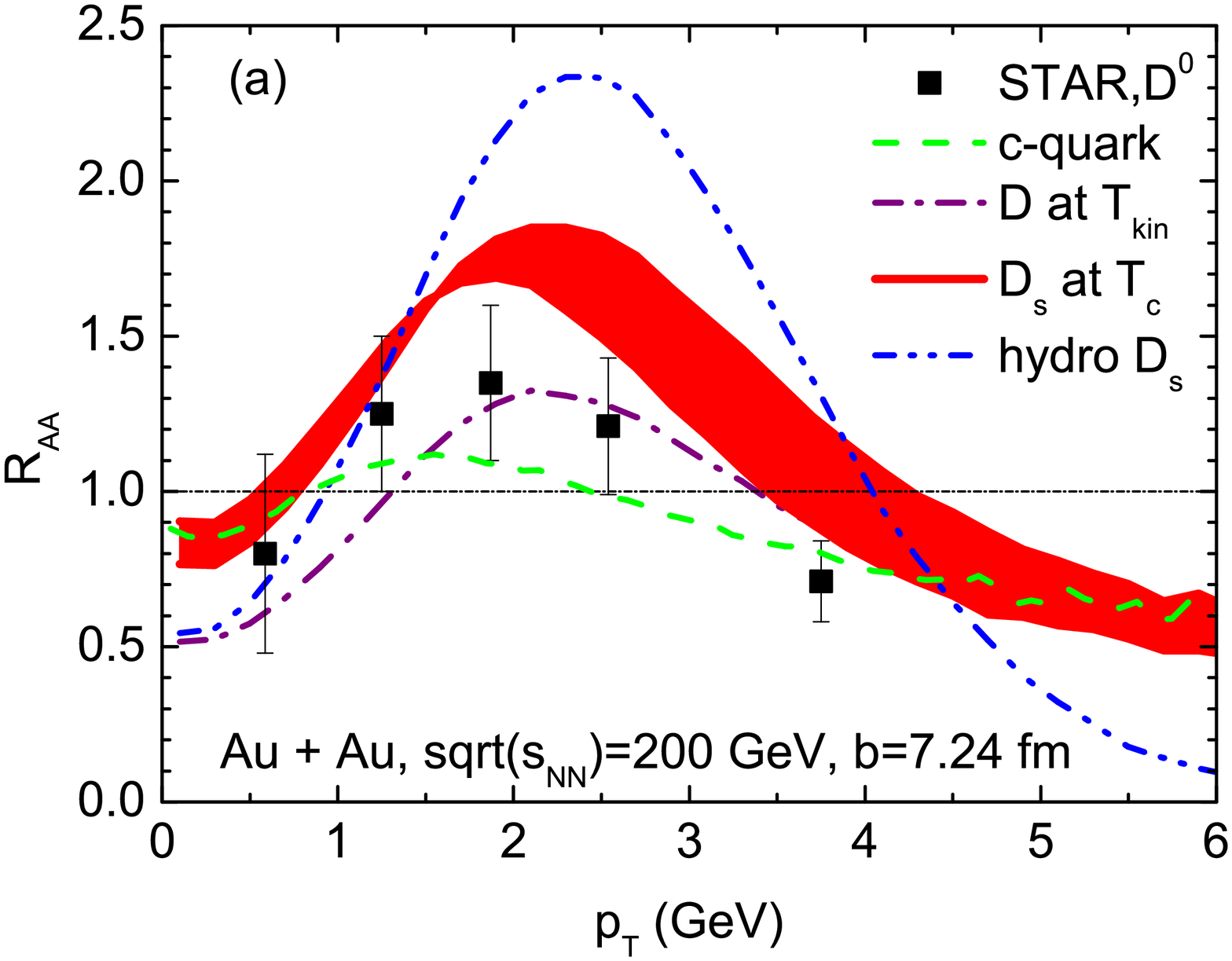}

\vspace{-0.6cm}
\includegraphics[width=\columnwidth]{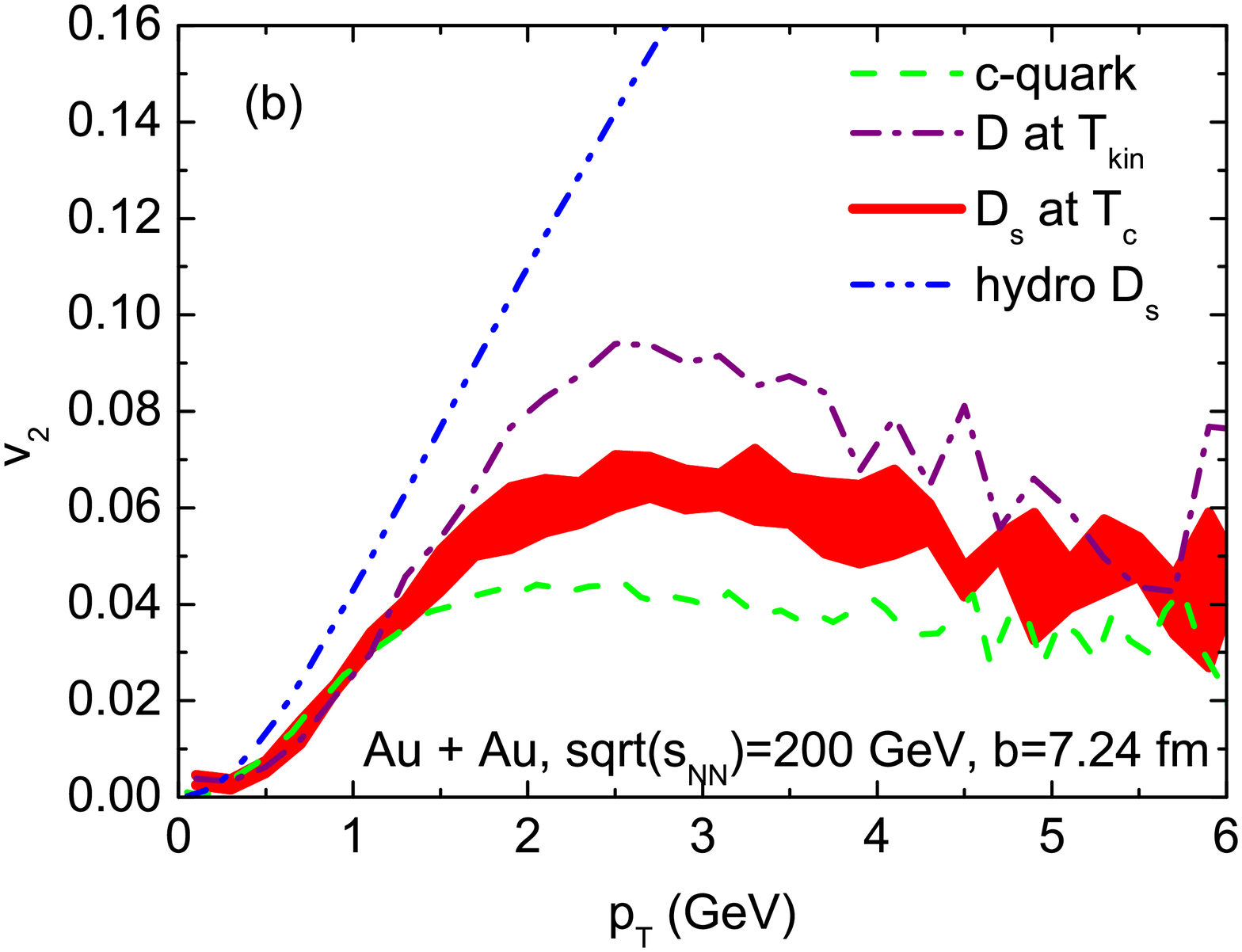}
\caption{(Color online) Our results for the nuclear
modification factor (upper panel) and elliptic flow (lower panel) of
$D_s$ (red bands) and $D$ mesons (green dash-dotted lines) in
semi-central Au+Au collisions at RHIC. We also show the result for charm
quarks at $T_c$ (purple dashed lines), the equilibrium limit for $D_s$
mesons in the hydrodynamic medium at $T_c$ (blue dash-double-dotted line)
and preliminary STAR data~\cite{Zhang:2011uva} for the $D$-meson $R_{\rm AA}$ in
0-80\% Au+Au. In the upper panel, the red uncertainty band is governed
by inclusion or omission of a Cronin effect in the initial charm spectra,
while in the lower panel it is controlled by including or neglecting diffusion
effects in the hadronization window.}
\label{fig_charmDDs}
\end{figure}

Figure~\ref{fig_charmDDs} summarizes our $D$ and $D_s$ meson spectra
in semi-central Au+Au at RHIC relative to $pp$ collisions in
terms of the nuclear modification factor,
$R_{AA}(p_T)=(dN^{AA}/dp_T)/(N_{\rm coll}dN^{pp}/dp_T)$ ($N_{\rm coll}$:
number of binary $NN$ collisions in AA), and elliptic-flow
coefficient, $v_2(p_T)$ (the second harmonic of the azimuthal-angle
dependence).
Both the $D$ and $D_s$ $R_{AA}$ (upper panel) exhibit a maximum around
$p_T$$\simeq$2-3\,GeV, induced by the transverse flow picked up from
the expanding medium. Current STAR data are consistent with our $D$-meson
result, but we predict the maximum to be more pronounced for the $D_s$
(reaching beyond 1.5) due to $c$-quark coalescence with the enhanced
strangeness in Au+Au.
To further illustrate this effect, we also plot the result for $c$-quarks at
the end of the QGP phase, which would directly represent $D$-
and $D_s$-spectra if coalescence were absent and only $\delta$-function
fragmentation applied (as in $pp$). One clearly recognizes the
important effect of coalescence, which only ceases
above $p_T$$\simeq$5\,GeV, where fragmentation takes over and the $D$, $D_s$ and $c$-quark $R_{AA}$ merge.
While the $c$-quark spectra are not observable, the $D$ and $D_s$ ones
are, so that their difference gives a quantitative measure of
the coalescence effect. It turns out that hadronic diffusion does
not significantly affect the $D$-meson $R_{AA}$ (due to a compensation
of a decreasing temperature and an increasing flow of the medium).

The elliptic flow of particle spectra is known to be an excellent measure
of the medium's collectivity due to hydrodynamic flow in non-central AA
collisions (induced by the ``almond-shaped" initial nuclear overlap zone).
In our calculations, the diffusion in the QGP imparts an appreciable
$v_2$ on the charm quarks of up to $\sim$4.5\%, cf.~lower panel
in Fig.~\ref{fig_charmDDs}. Coalescence with thermal quarks amplifies
this value by up to 50\%, for both $D$ and $D_s$ mesons.
However, while the $D_s$ spectra freeze out after hadronization,
the $D$ coupling to the hadronic medium, which inherits the full
elliptic flow from the QGP expansion~\cite{He:2011zx}, further augments
$v_2$ by up to 30\%. We therefore suggest the $v_2$-splitting
between $D$ and $D_s$, in the spirit of the early multistrange freezeout
in the underlying hydro evolution,  as a promising measure of
the transport properties of the hadronic phase.

In summary, we have argued that measurements of $D$ vs.~$D_s$-meson
$R_{AA}$ and $v_2$ in URHICs provide powerful constraints on
heavy-flavor diffusion and hadronization. We have made predictions for
these observables employing a self-consistent framework where the
concept of a strongly coupled QGP is implemented in both macro- and
microscopic components of the calculation:
a hydrodynamic medium evolution, quantitatively tuned to bulk- and
multistrange-hadron observables, has been combined with nonperturbative
charm-diffusion coefficients in the QGP which are compatible with
currently available lQCD results.
The diffusion of $D$ mesons in the hadronic phase has been implemented for
the first time, while $D_s$ mesons are frozen out at $T_c$.
A remarkable enhancement of the $D_s$-meson $R_{AA}$ well above one
emerges as a result of a strong charm-quark coupling to the QGP and
subsequent recombination with equilibrated strange quarks. The
latter can be directly tested by comparing $D_s$- and $D$-meson $R_{AA}$.
The $D$ meson picks up significant additional $v_2$ from the hadronic
phase, which can be quantified by comparing to $D_s$-meson $v_2$
for which hadronic diffusion effects are absent. This picture should
persist at LHC and can be directly carried over to the bottom sector
using $B$ and $B_s$ mesons.

{\it Acknowledgments.---}
This work was supported by the U.S. National Science Foundation
(NSF) through CAREER grant PHY-0847538 and grant PHY-0969394, by the
A.-v.-Humboldt Foundation, and by the JET Collaboration and DOE grant
DE-FG02-10ER41682.

\end{document}